\documentstyle[12pt,epsfig]{article}
\newlength{\dinwidth}
\newlength{\dinmargin}
\newcommand{\ba}{\begin{array}}
\newcommand{\ea}{\end{array}}
\newcommand{\be}{\begin{equation}}
\newcommand{\ee}{\end{equation}}
\newcommand{\bea}{\begin{eqnarray}}
\newcommand{\eea}{\end{eqnarray}}

\newcommand{\nono}{\nonumber}
\newcommand{\Ac}{{\cal A}_c}
\newcommand{\Ao}{{\cal A}_o}

\newcommand{\Ast}{{\cal A}_{st}}
\newcommand{\Ec}{{\cal E}_c}
\newcommand{\Eo}{{\cal E}_o}
\newcommand{\del}{\partial}
\newcommand{\M}{s\ln |s|+t\ln |t|+u\ln |u|}
\newcommand{\G}{\Gamma}

\newcommand{\gsim}{\mathrel{\mathop{\kern 0pt \rlap
  {\raise.2ex\hbox{$>$}}}
  \lower.9ex\hbox{\kern-.190em $\sim$}}}

\begin{document}
\thispagestyle{empty}
\addtocounter{page}{-1}
\begin{flushright}
SNUST-000601\\
{\tt hep-th/0007055}
\end{flushright}
\vspace*{1.3cm}
\centerline{\Large \bf Time-Delay at Higher Genus}
\vspace*{0.4cm}
\centerline{\Large \bf in}
\vspace*{0.4cm}
\centerline{\Large \bf High-Energy Open String Scattering~\footnote{
Work supported in part by BK-21 Initiative in Physics (SNU - 
Project 2), KRF International Collaboration Grant 1998-010-192, KOSEF 
Interdisciplinary Research Grant 98-07-02-07-01-5, and KOSEF
Leading Scientist Program 2000-1-11200-001-1.}}
\vspace*{1.2cm} 
\centerline{\bf Tsunehide Kuroki {\rm and} Soo-Jong Rey}
\vspace*{0.8cm}
\centerline{\it School of Physics \& Center for Theoretical Physics}
\vspace*{0.3cm}
\centerline{\it Seoul National University, Seoul 151-742 Korea}
\vspace*{1cm}
\centerline{\tt kuroki@phya.snu.ac.kr \hskip1cm sjrey@gravity.snu.ac.kr}
\vspace*{1.5cm}
\centerline{\bf abstract}
\vspace*{0.5cm}
\noindent
We explore some aspects of causal time-delay in open string scattering
studied recently by Seiberg, Susskind and Toumbas. By examining high-energy 
scattering amplitudes at higher order in perturbation theory, we argue that 
causal time-delay at $G$-th order is $1/(G+1)$ times smaller than the 
time-delay at tree level. We propose a space-time interpretation of the result 
by utilizing the picture of the high-energy open string scattering put forward 
by Gross and Ma\~{n}es. We argue that the phenomenon of reduced time-delay is 
attributed to the universal feature of the space-time string trajectory in 
high-energy scattering that string shape at higher order remains the same as 
that at tree level but overall scale is reduced. 
We also discuss implications to the space-time uncertainty principle 
and make brief comments on causal time-delay behavior in space/time 
noncommutative field theory.
\vspace*{1.1cm}

\baselineskip=18pt
\newpage

\section{Introduction}
Noncommutative structures in string theory seem worth studying 
not only because they arise as a nonperturbative effect as in 
D-branes \cite{witten,DH,ishi,KK,SW}, Matrix theory \cite{BFSS,CDS}, 
or IIB matrix models \cite{IKKT,AIKKT}, but because they would reveal 
essential features of string theory as a unique theory known to date 
describing various extended objects (including strings themselves) 
in a consistent manner.  However, recent progress in the study of 
these structures suggests that the extendedness of a string in the space-time 
cannot be described by a mere specification of noncommutativity relation 
such as 
\be
[X^{\mu}, X^{\nu}]=i\theta^{\mu\nu} 
\label{nc}
\ee
but requires more specification concerning microscopic details.
In fact, quantum field theory on a noncommutative space defined by 
the above commutation relation with $\theta^{0i} \neq 0$, where $i$ denotes 
one of the spatial directions, is found to behave as a theory consisting of 
extended objects like dipoles \cite {SST}. Consequently, from the viewpoint 
of pointlike quanta, the theory exhibits acausal behavior, nonlocality and 
nonunitarity \cite{SST, gomis}. 
On the other hand, it is observed in \cite{SST, sjrey} that open strings living
on the same noncommutative space do not show such pathologies 
and exhibit causal time-delays, reflecting their extendedness. 
Furthermore, a space/time noncommutative field theory cannot be obtained 
as the low-energy limit of string theory \cite{SST2, GMMS, BR, reyvonunge}. 
Thus the open string exhibits another origin of its consistent extendedness 
other than noncommutativity described by (\ref{nc}). 
{}From this point of view, it is important to examine causal time-delay
behavior of the open string theory in further details 
and check its consistency. Another importance is in that it seems related to 
the space-time uncertainty principle \cite{stur, yoneya}. 
This is as it should be, 
because the space-time uncertainty principle basically prescribes 
how a string extends in space-time in a way consistent with (perturbative)
unitarity and analyticity. 
Therefore, studying the causal time-delay behavior may shed light 
on identifying central features in string theory dynamics.
 
In the present paper, we shall be exploring causal time-delay that show up
in high-energy scattering of open strings at fixed order in string
 perturbation theory. Our motivation for the study comes from various
sides. The high-energy scattering may enable us to explore 
the short-distance structure of string theory. 
The reason why we go beyond the tree-level analysis done in \cite{SST} 
is that in the high-energy scattering the contribution from higher genus 
to the amplitude is known to dominate the lower order ones 
\cite{GM1, GM2, GMan}. 
Therefore, it is important to examine the causal time-delay at higher
orders and see if it is pronounced similarly as compared to the result of 
\cite{SST}. 
Another motivation is that at high energy the space-time picture 
of the scattering becomes transparent because a certain class of classical 
trajectories become dominant \cite{GM1, GM2, GMan}. 
This effect permits examining whether the interpretation of time-delays 
given in \cite{SST} persists to hold at higher genus as well.  

The organization of this paper is as follows. In the next section, 
we give a brief recapitulation of the high-energy open string scattering 
at an arbitrary order in perturbation theory \cite{GMan}. 
In section 3 we proceed to evaluate the causal time-delay at each order in 
perturbation theory and find that it is reduced gradually at higher orders. 
In section 4, we discuss space-time interpretation of our result 
using various features of the saddle point configuration \cite{GMan}. 
Our analysis 
shows that the result is consistent with the interpretation in \cite{SST} 
based on the property that the string grows in the longitudinal 
direction with its energy. 
In the final section, we make brief comments 
on relation with the space-time uncertainty principle \cite{stur, yoneya} 
and comparison with the case of space/time noncommutative field theory.  


\section{High-Energy Scattering of Open Strings}
\setcounter{equation}{0}
In this section, for our foregoing analysis, we recapitulate relevant 
results concerning the high-energy scattering of open strings \cite{GMan}. 
Consider, at genus $G$, four-point scattering amplitude in {\sl closed }
string theory given by:
\be
\Ac^{(G)}(p_1, \cdots, p_4) =g^{2G+2}\int\frac{{\cal D}g_{\alpha\beta}}
                           {{\cal N}}{\cal D}X^{\mu}
                      \exp\left(\frac{1}{4\pi\alpha'}
                              \int d^2\xi\sqrt{g}g^{\alpha\beta}
                              \del_{\alpha}X^{\mu}\del_{\beta}X_{\mu}
                          \right)\Pi_{i=1}^{4}V_{i}(p_i),    
\label{closedamp}
\ee
where $g$ is the closed string coupling constant, $\cal{N}$ is the volume 
of the group of the diffeomorphisms and Weyl rescalings, and $V_i(p_i)$ are 
the vertex operators. 
In the case of the scattering of four tachyons, the Gaussian integration 
over $X^{\mu}$ yields 
\be
\Ac^{(G)}(p_i)=g^{2G+2}\int_{{\cal M}_G}[dm]\Pi_id^2\xi_i\sqrt{g(\xi_i)}
               \Omega_c(m,\xi_i)
               \exp\left(-\alpha'\sum_{i<j}p_i\cdot p_j~
                                           G_m(\xi_i,\xi_j)
                   \right),
\nonumber
\ee
where $m$ are a set of coordinates for the moduli space ${\cal M}_G$ 
of genus $G$ Riemann surfaces with four punctures, 
$\Omega_c(m,\xi_i)$ is the standard measure on ${\cal M}_G$, and $G_m$ is the 
scalar Green function on the genus-$G$ Riemann surface.  
In the kinematical limit all $p_i\cdot p_j$'s become large, one can perform 
the integration via saddle-point method and get \cite{GM1, GM2} 
\be
\Ac^{(G)}(p_i)\sim g^{2G+2}\Omega_c(\hat{m},\hat{\xi}_i)
                   ({\cal E}_c'')^{-\frac{1}{2}}
                   \exp\left(-\alpha'\Ec (p_i,\hat{\xi}_i,\hat{m}_i)\right),
\nonumber
\ee
where 
\be
\Ec(p_i,\xi_i,m)\equiv\sum_{i<j}p_i\cdot p_j~G_m(\xi_i,\xi_j)
\nonumber
\ee
is the electrostatic energy of an analog system of charges $p_i$ at $\xi_i$ 
on a Riemann surface with moduli $m$. $\hat{\xi}_i$ and $\hat{m}$ denote 
the values at the saddle point where $({\cal E}_c'')^{-1/2}$, the determinant 
of the second derivative of $\Ec$, is evaluated.  

Following the strategy of \cite{GMan}, high-energy scattering amplitudes of
open strings may be obtained from those of closed strings by utilizing 
the reflection principle: doubling an open Riemann surface with boundaries 
yields a closed Riemann surface of higher genus. 
It turns out that, in the oriented open string theory, 
all relevant quantities associated with the original open Riemann surface 
can be extracted from the doubled, 
closed Riemann surface by imposing invariance 
under the reflection with respect to an appropriate symmetry plane \cite{GMan}. 
Since (\ref{closedamp}) can be easily generalized to Riemann surfaces 
with boundaries, the above argument shows that, in the high energy 
scattering of four open strings as well, the amplitude are approximated by 
its saddle point expression at the $G$-th order in open string 
perturbation theory 
\be
\Ao^{(G)}(p_i)\sim g^{G+1}\Omega_o(\hat{m},\hat{\xi}_i)
                   ({\cal E}_o'')^{-\frac{1}{2}}
                   \exp\left(-\alpha'\Eo (p_i,\hat{\xi}_i,\hat{m}_i)\right),
\nonumber
\ee
where $\Omega_o$ denotes the measure on the moduli space of open Riemann
surfaces with boundaries,
\be
\Eo(p_i,\xi_i,m)\equiv 2\sum_{i<j}p_i\cdot p_j~G_m(\xi_i,\xi_j),
\nonumber
\ee
is the analog electrostatic energy on the open Riemann surface, while
 $G_m$ denotes the scalar Green function for the doubled, closed 
Riemann surface of genus $G$. 
In fact, it can be shown that 
every saddle-point configuration of the oriented open string 
can be constructed from the associated saddle-point configuration 
of the closed string via reflection principle \cite{GMan}. 
The closed string saddle-point configuration at the $G$-th order was 
obtained in \cite{GM1, GM2} as the $(G+1)$-sheeted Riemann surface of the form 
\be
y^{G+1}=\Pi_{i=1}^4(z-a_i)^{L_i}, 
\label{Nsheet}
\ee
where $L_i$'s are relatively prime to $G+1$, 
$\sum_iL_i=0$ (mod $G+1$)\footnote
{Note that this condition prevents from considering all possible 
Riemann surfaces. Just for brevity, we will not treat the case of 
$L_i\neq 0$ (mod $G+1$).} 
and the branch points $a_i$ are separated by $1/(G+1)$ times the period. 
Therefore one can construct the corresponding open string saddle-points 
and also evaluate the electrostatic energy $\Eo^{(G)}$ on them. 
It is given by 
\be
\Eo^{(G)}=\frac{\Eo^{(0)}}{G+1},
\nonumber
\ee
where 
\be
\Eo^{(0)}=\M,
\nonumber
\ee
is the electrostatic energy on the disk, and $s$, $t$, $u$ are the Mandelstam 
variables: $s=-(p_1+p_2)^2$, $t=-(p_2+p_3)^2$,  $u=-(p_1+p_3)^2$. 
Thus, the high-energy scattering amplitude of four open tachyons at the 
$G$-th order in perturbation theory is approximated by 
\be
\Ao^{(G)}\sim g^{G+1}\exp\left(-\alpha'\frac{\M}{G+1}\right).  
\label{GManamp}
\ee

{}For later convenience, let us consider the nearly backward scattering 
$u\sim 0$ and compare the amplitude Eq.(\ref{GManamp}) at $G=0$ with the 
Veneziano amplitude in the same kinematical region. The Veneziano amplitude 
is given by 
\be
\Ast\sim g\frac{\G (-\alpha's)\G (-\alpha't)}{\G (1+\alpha'u)}
       +(t\leftrightarrow u)
       +(s\leftrightarrow u),
\label{veneziano}
\ee
up to kinematical factors. For small $u$, the first term can be expanded as 
\be
\Ast\sim g\frac{\pi}{s\sin (\pi\alpha's)}(1-\alpha'u\ln(\alpha's)+\cdots)
\label{nearbackexp}, 
\ee
where we use the identity
\be
z\G (-z)\G (z)=\frac{-\pi}{\sin \pi z}.
\nonumber
\ee
It should be noted that as long as $u$ is finite {\sl and} non-zero, 
the second term in (\ref{nearbackexp}) is dominant in the high energy region, 
while setting $u$ to zero, only the first term in (\ref{nearbackexp}) 
would survive. This manifests the fact that the high-energy
limit $s\rightarrow\infty$ does not commute with the backward scattering 
limit $u\rightarrow 0$. Anther way to see this 
is that the high-energy scattering amplitude (\ref{GManamp}) becomes trivial 
if one sets $u=0$, a result which does not agree with (\ref{nearbackexp}). 
In fact, the above analysis cannot be applied to the exactly forward ($t=0$) 
or exactly backward ($u=0$) scattering, as, in such cases, the saddle
point configurations approach 
boundaries of the moduli space ${\cal M}_G$, where two of the points $\xi_i$ 
at which vertex operators are inserted become coincident. 

To proceed further and examine causal time-delay, we will need to have
at hand asymptotic scattering amplitudes which are valid even 
at higher perturbative order. 
Therefore, in the following, we shall consider the high-energy, fixed angle 
scattering with a moderate scattering angle $\phi\neq 0, \pi$, 
where we can rely on the Gross-Ma\~{n}es amplitude (\ref{GManamp}). 
Note that this is in contrast to the case in \cite{SST} 
where the exactly backward scattering is considered and accordingly 
the scattering amplitude becomes singular in the sense that all poles 
contribute to it. 

In concluding this section, it is worth pointing out a universal structure 
of the Gross-Ma\~{n}es amplitude (\ref{GManamp}).  
At the $G$-th order in string perturbation theory, 
the $G$-dependence in (\ref{GManamp}) prompts 
to perform the following rescaling:
\be
\alpha'\rightarrow\hat{\alpha}'\equiv\frac{\alpha'}{G+1}.
\label{scaling}
\ee
Defining rescaled kinematic invariants
\bea
\hat{s}\alpha' & \equiv & s\hat{\alpha}'=\frac{s}{G+1}\alpha', \nono \\
\hat{t}\alpha' & \equiv & t\hat{\alpha}'=\frac{t}{G+1}\alpha', \nono \\
\nonumber
\eea 
it follows that  
\be
\Ao^{(G)}(s,t)
\sim g^G \Ao^{(0)}(\hat{s},\hat{t}),
\label{Gamp} 
\ee
viz. the $G$-th order high-energy scattering amplitude can be reinterpreted
as the tree-level amplitude at reduced values of kinematical invariants.
As we will see later, universal features of the saddle 
point configurations yield a general and reasonable physical interpretation 
to the scaling (\ref{scaling}).


\section{Causal Time-Delay at Higher Genus}
\setcounter{equation}{0}
In this section, we would like to examine causal time-delay in the 
higher-order, high-energy scattering of four massless open string states. 
In order to do this, we first note that the saddle-point trajectory is 
universal and independent of the quantum numbers of scattered particles 
as the contribution from the vertex operators in (\ref{closedamp}) 
is at most the polynomials of momenta \cite{GM1, GM2, GMan, gross}. 
Thus, up to kinematical prefactors, the tachyon amplitude
(\ref{GManamp}) should be applicable to the massless case. 

Let us consider the scattering of two massless open string states 
$1+2 \rightarrow 3+4$, closely following the analysis of \cite{yoneya}. 
For simplicity, we assume that the scattering takes place 
in a plane in a flat space-time and choose the center-of-mass system 
for the transverse momenta. 
As suggested in \cite{yoneya}, one of the conceivable ways at present 
to extract the extendedness of strings from a scattering amplitude 
is that we first treat initial and final string states just 
as particle states by constructing their wave-packets with respect to 
the center-of-mass coordinates, and then observe the uncertainties 
of the interaction region, on which the extendedness would be reflected. 

Following this method, 
we choose the wave-packet for each massless open string states as 
\be
\Phi_i(x_i,p_i)=\int d{\bf k}_i f_i({\bf k}_i-{\bf p}_i) 
\exp\left(i({\bf k}_i\cdot{\bf x}_i-|{\bf k}_i|t_i)\right),  
\label{wavepacket}
\ee
where $f_i({\bf k})$ is any function with a peak at ${\bf k}=0$. 
Then the S-matrix element is given by 
\be
\langle 3,4|S|1,2\rangle = \left(\Pi_{i=1}^{4}\int d{\bf k}_i\right)
                           f_3^*f_4^*f_1f_2~\delta(\sum_{i=1}^{4}k_i)
                           {\cal A}(s,t).
\label{smatrix}
\ee
As proposed in \cite{yoneya}, the uncertainty of the interaction region 
is measured by examining the response of the S-matrix under appropriate 
shifts of the particle trajectories in space-time. If we make shifts 
$t_i\rightarrow t_i+\Delta t_i$, then the wave-packet 
(\ref{wavepacket}) becomes 
\be
\Phi_i(x_i,p_i;\Delta t_i)=\int d{\bf k}_i f_i({\bf k}_i-{\bf p}_i) 
\exp\left(i({\bf k}_i\cdot{\bf x}_i-|{\bf k}_i|t_i)\right) 
\exp(-i|{\bf k}_i|\Delta t_i).  
\ee
For simplicity, set 
$\Delta t_1=\Delta t_2=-\Delta t_3=-\Delta t_4=\Delta t/2$ 
and $|{\bf k}_i|=E$ for all $i$, the integrand in (\ref{smatrix}) 
would acquire an additional phase factor $\exp(-2iE\Delta t)$ 
due to this shift. 
Then the time-delay $\Delta T$ (uncertainty in time) can be estimated 
as the decay width of (\ref{smatrix}) with respect to $|\Delta t|$ 
under the insertion of this additional phase factor. 
As argued in \cite{yoneya}, as $\Delta t$ increases, the decay becomes 
appreciable when the variation of $\ln{\cal A}(E)$ is exceeded 
by the variation of the additional phase $2E\Delta t$. 
Therefore, we can set 
\be
\Delta T \sim \langle |\Delta t| \rangle \sim 
              \frac{1}{2}\left|\frac{\del}{\del E}\ln {\cal A}(E)\right|.
\label{delay}
\ee
Substituting the Gross-Ma\~{n}es amplitude (\ref{GManamp}) 
into this equation and setting $E=|{\bf p}|=p_0$, we immediately obtain 
\be
\Delta T \sim \frac{\alpha' p_0}{G+1}f(\phi),
\label{reduceddelay}
\ee
where $f(\phi)$ is defined as 
\be
f(\phi)\equiv -\sin^2\frac{\phi}{2}\ln\sin^2\frac{\phi}{2}
              -\cos^2\frac{\phi}{2}\ln\cos^2\frac{\phi}{2},
\ee
thus fixed in our kinematical region. 
Although we cannot determine the sign of the time uncertainty by this method, 
we conclude that this is causal, because at tree level it is argued 
in \cite{SST, BR} that the string scattering is indeed causal, 
and even at higher genus, the scattering process remains almost the same 
as that at tree level due to universal features of the Gross-Ma\~{n}es saddle 
point, as we will discuss in the next section.   
The reduced causal time-delay (\ref{reduceddelay}) is the main result 
in this section.
It displays the fact that the causal time-delay at $G$-th order in string
perturbation theory is reduced by $1/(G+1)$ compared to that at tree level. 
Here we note that, in (\ref{delay}), a numerical constant 
we have omitted should be independent of the genus, because it originates 
only from kinematics of the scattering process. Rather, it may be possible 
to take (\ref{delay}) as one of definitions of $\Delta T$ in the framework 
here. 

It should be emphasized that the time-delay at each order in perturbation 
theory is purely theoretical: what we will observe here must be a time-delay 
associated with the full amplitude. Nevertheless we believe our reduced time 
delay is important because, as we discussed later, it provides reasonable 
physical pictures and important implications of high-energy behavior 
of strings in the framework of perturbative string theory.   

\section{Reduction of Time-Delay: Space-Time Interpretation}
\setcounter{equation}{0}
In this section, we would like to draw certain space-time interpretation 
concerning the higher-genus reduction of the causal time-delay 
(\ref{reduceddelay}) by collecting various features the saddle-point
exhibits. 

Let us begin with the saddle-point trajectory 
of the closed string as originally found in \cite{GM2}. It is given by 
\be
X^{\mu}(z)=\frac{i}{G+1}\sum_{i=1}^{4}\alpha'p_i^{\mu}\ln |z-a_i|
          +{\cal O}(1/s), 
\label{saddle}
\ee
where $z$ denotes the point on the $(G+1)$-sheeted Riemann surface 
(\ref{Nsheet}). It is worth noting that, from (\ref{saddle}), even at 
higher order, the shape of the saddle point trajectory remains the same as 
that at tree level, except that the overall {\sl space-time} scale is
reduced by a factor $1/(G+1)$. By examining the behavior at the vicinity 
of the branch points $a_i$, where the four vertex operators are inserted, 
one can easily get the following space-time picture of the saddle point
trajectory \cite{GM2}: each of 
the two incoming closed strings winds around a closed curve $(G+1)$ times, 
then they interact and propagate as $(G+1)$ many 
intermediate short closed strings.\footnote
{To be more precise, there exist the cases in which we have only one 
intermediate string. They occur when $L_i+L_j\neq 0$ (mod $G+1$) 
in (\ref{Nsheet}). However, they does not make a significant difference 
in later discussions. See also the footnote below Eq. (\ref{Nsheet}).} 
Subsequently, the $G+1$ short strings then rejoin together and finally
produce two separate $(G+1)$-times wound outgoing closed strings. 
This picture is consistent with the fact that, as shown in (\ref{saddle}),
the string trajectory at 
$G$-th order is scaled down by a factor $1/(G+1)$ compared to that 
at tree level. 

As recapitulated in the previous section, any oriented open string diagram can 
be obtained from corresponding closed string diagram by cutting each closed 
string into two open strings and keep one side of them. Therefore, high-energy 
scattering amplitude and space-time picture for open string can be obtained, 
via reflection principle, from those for closed string. From the space-time 
picture of the open string trajectory deduced this way, it follows immediately 
that, for the open strings as well, the space-time
trajectory of the saddle point configuration at higher order in perturbation 
theory is exactly the same as that at tree level, except now that the
space-time size is reduced by a factor $1/(G+1)$. 

In interpreting the causal time-delay of open string scattering 
at tree level \cite{SST}, key observations have been that the extension of 
the open string along the longitudinal direction 
grows linearly with energy, $L\sim p_0\alpha'$, and that this leads to 
causal time-delay whose magnitude is proportional to $L$ and hence to 
$p_0$.\footnote
{Here, one assumes implicitly that the longitudinal direction 
of a string could be identified definitely. However, it is not evident that 
this is always possible especially at high energy where a string 
would oscillate frequently. Presumably our scattering should be 
argued in the infinite momentum frame which is the most natural in 
high-energy scattering.} 
{}From this point of view, the fact that saddle-point trajectories
of the open string are all universal enables us to understand 
intuitively  reduction of the causal time-delay (\ref{reduceddelay}) 
at higher orders. At $G$-th order in perturbation theory, 
size of each open string along the longitudinal direction (scattering
axis) is effectively scaled down by $1/(G+1)$ compared to that at tree level. 
Consequently, when scattered off each other, the open strings at $G$-th order
would display the causal time-delay only by $1/(G+1)$ times 
that at tree level. 
This intuitive explanation is in complete agreement with our
earlier result (\ref{reduceddelay}). 
Stated differently, as the incoming strings 
are folded or wound around $(G+1)$ times, their tension is effectively 
increased by a factor $(G+1)$ as in (\ref{scaling}).\footnote
{As the 
string is folded over, one might have anticipated that the tension is 
reduced instead of being increased. However, this argument applies only
to nonrelativistic string such as polymer. For relativistic string, the
tension remains always equal to the mass density.} Correspondingly, 
the causal time-delay is reduced as given in (\ref{reduceddelay}). 
Turning around the argument, from high-energy scattering in perturbation 
theory, we have confirmed that the causal time-delay exhibited therein is 
consistent with the interpretation that the longitudinal size of the 
string grows linearly with energy, $L\sim p_0\alpha'$ (see (\ref{saddle})). 
This seems also consistent with the space-time uncertainty principle 
\cite{stur, yoneya}, on which we will discuss more later.    

\begin{figure}[t]
   \vspace{1cm}
   \epsfysize=8cm
   \epsfxsize=13cm
   \centerline{\epsffile{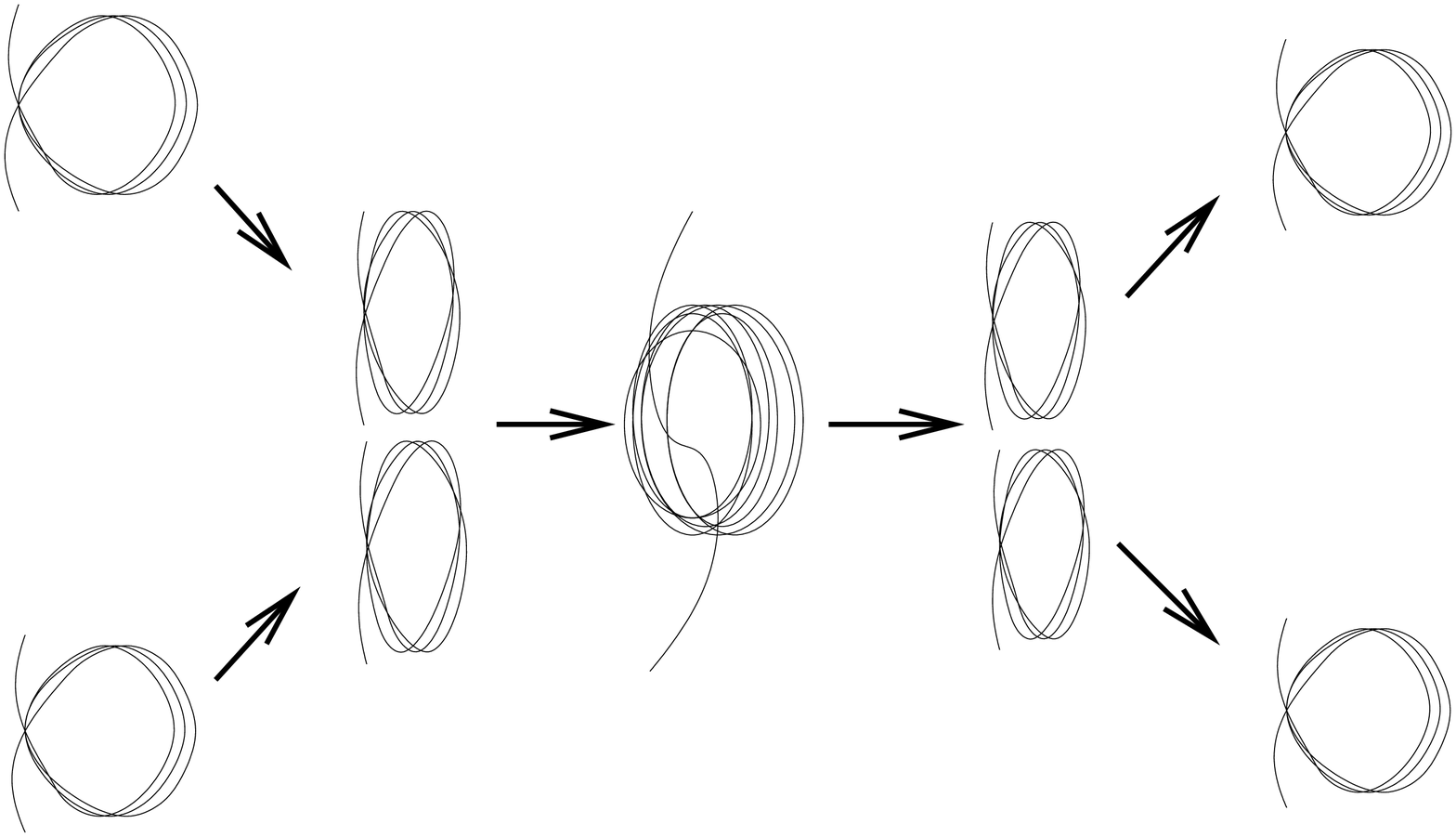}}
   \vspace*{.5cm}
\caption{\label{fig1} Space-time picture of the high-energy open string
scattering at $G$-th order in perturbation theory.} 
\end{figure}

In concluding this section, we summarize 
the space-time trajectory of high-energy open string
scattering \cite{GMan} as follows: according to 
(\ref{saddle}) and the reflection principle, each of the two incoming open
strings is in $(G+1)$-times multiply wound or folded configuration. As the
two open string endpoints interact, at the scattering point, the wound or 
folded strings rearrange themselves into a configuration consisting of
at most two short open strings and $\sim G/2$ many singly wound, short closed 
strings. Characteristic size of these short strings is that of incoming
open strings, viz. $1/(G+1)$-times smaller than the size at tree level.
After interaction, the little strings rejoin and split off into two 
outgoing open strings, each of them are again $(G+1)$-times multiply wound
or folded. In general, these outgoing open strings would have different 
lengths from the incoming ones. A cartoon view of the space-time trajectory
of the scattering process is depicted in Figure 1. 


\section{Discussions}
\setcounter{equation}{0}
\subsection{Comparison with Space/Time Noncommutative Field Theory}
Let us compare our results with those expected in the space/time 
noncommutative field theory. As shown in \cite{SST}, at tree level, the
noncommutative field theory exhibits both advanced and retarded scattering. 
Even though the theory is generically non-unitary \cite{gomis}, 
one might put it aside
and inquire of acausality at higher orders in perturbation theory:
are the advanced/retarded effects present at higher orders in perturbation
theory and, if so, are the effects increased or reduced as compared
to those at tree level? It is well known that, in the maximal noncommutativity 
limit, $\theta^{\mu \nu}\rightarrow\infty$, the scattering amplitudes of 
noncommutative field theory are dominated by planar loop graphs. This is 
so because all nonplanar loop graphs are completely suppressed due to the 
maximal noncommutativity. In planar graphs, there is no $\theta$-dependence 
apart 
from the overall Moyal phase associated with external momenta. These 
overall Moyal phase is the same for all planar graphs no matter how higher
order they come from in perturbation theory. As such, at 
any order in perturbation theory, one does not expect any reduction of the 
retarded or advanced scattering behavior in the maximally noncommutative 
field theory. This should be contrasted to 
the fact that, in open string theory, universal structure of the saddle 
points has played an essential role in the reduction of the causal time 
delay. Indeed, exponential fall off and universal structure of the 
high-energy scattering amplitude (\ref{GManamp}) are characteristic features 
of open string theory, which are not present in quantum field theories
with a finite number of field degrees of freedom.  

\subsection{Relation to Space-time Uncertainty Relation}
Let us make comparison of what we have found with the space-time 
uncertainty relation:
\cite{stur, yoneya}
\be
\Delta X \Delta T \gsim \alpha'.
\label{stup}
\ee
First it is worth noting that in our case this inequality is far from 
being saturated at lower order in perturbation theory. One might naively
estimate that, in the present context, 
\be
\Delta X\sim\frac{p_0\alpha'}{G+1} \qquad \quad {\rm and} \qquad \quad
\Delta T\sim\frac{1}{E}\sim\frac{1}{p_0},
\nonumber
\ee
such that the space-time uncertainty relation (\ref{stup}) is not obeyed. 
However, $\Delta T$ should include not only the quantum fluctuation
uncertainty but also the causal time-delay effect and, at high energy 
$p_0^2 \alpha' \gg 1$, the latter is the dominant effect: 
\be
\Delta T\sim\frac{p_0\alpha'}{G+1}.
\label{delayed}
\ee
Thus, at a fixed order in perturbation theory, the left-hand side of 
(\ref{stup}) is much bigger than $\alpha'$.\footnote
{We would like to thank T. Yoneya for pointing out this effect.} 
Even including time-delay effect, the fact 
that both $\Delta X$ and $\Delta T$ become smaller at higher order
as estimated above 
implies that the space-time uncertainty relation (\ref{stup}) 
would be saturated at the order $G_{\rm max} \sim \sqrt {p_0^2 \alpha'}$. 
Does this mean that the space-time uncertainty relation is violated beyond 
that order in perturbation theory? We believe not so.

A possible resolution is that the causal time-delay may be enhanced further
at higher-order in perturbation theory. Indeed, in case of high-energy
{\sl closed} string scattering, via Borel-resummation of
fixed order saddle-point results \cite{MO}, it has been already argued that 
high-energy asymptotic behavior could be significantly modified at 
nonperturbative level.  After Borel-resummation, the scattering amplitude 
turns out to behaves as 
\be
{\cal A}_{resum}(s)\sim e^{-\sqrt{s}}  
\label{Borel}
\nonumber
\ee
in the kinematical range $\log(1/g^2)\ll s \ll 1/g^{4/3}$ and exhibits 
much bigger amplitude than any fixed-order perturbative behavior, 
$e^{-s/(G+1)}$. Thus, once Borel-resummed, 
it might be that causal time-delay is considerably different 
from the fixed-order estimate (\ref{delayed}). 
It would even be the case that the inequality in the space-time uncertainty
relation (\ref{stup}) is saturated nonperturbatively in the 
high-energy regime, where the symmetry of the string theory is also 
believed to be enhanced \cite{gross} enormously. 
In \cite{yoneya}, the time-delay is indeed discussed based on (\ref{Borel}). 
However, it should be noted that (\ref{Borel}) was derived by the Borel 
resummation and as such it is one of possible guesses for nonperturbative 
high-energy amplitude. On the other hand, our time-delay is based on 
the amplitude at the saddle point which is certainly valid in the 
high-energy scattering with a fixed moderate angle. In this sense, 
we believe that although our time-delay is theoretical in itself, 
its reduced nature is important and that it indeed suggests 
the nonperturbative saturation of the space-time uncertainty relation. 
This is because if it were to grow with the order of the perturbation theory, 
it is impossible to expect such saturation. 
One can also find another importance of our time-delay in the following 
heuristic argument: suppose it is also possible to take advantage of 
the saddle point method in the summation over $G$ as discussed in \cite{GM1}, 
we find that the dominant order is given by $G \sim \sqrt{-\alpha' s/\log g}$, 
and then we get $\Delta X \sim \Delta T \sim \sqrt{\alpha'}$ up to 
some logarithmic corrections by substituting this into our time-delay.  
Therefore, this argument using our time-delay also implies that 
the space-time uncertainty relation is almost saturated at nonperturbative 
level. 

Another possible resolution is that yet another new sort of nonperturbative 
effects
begin to appear at the order $G_{\rm max}$. For example, as the strings are
boosted to infinite momenta, $p_0 \sim p_\parallel \sim N/R$ ( $R$ is
an appropriate infrared cutoff scale) and hence $G_{\rm 
max} \sim N$. This is the same order as the light-cone momenta $p_+$ and
hence the width of the light-cone string diagram. We have seen, in previous
sections, that the saddle-point trajectory is such that the string folds
or winds around $G$-times. The fact that there is a maximal folding or
winding and that each elementary folding or winding carries ${\cal O}(1)$ 
unit of
$p_+$ momentum indicate that sub-structure of the open string begins
to show up. Indeed, the effect is strikingly reminiscent of the matrix 
string theory, where a generic string configuration is built out of 
minimal length strings, each carrying precisely one unit of $p_+$. Using
matrix string theory, it has even been found \cite{GHV} that D-brane 
pair-production 
becomes an important effect when the fragmented strings produced
during the high-energy scattering are all of minimal length, viz. 
$G \sim G_{\rm max}$. Whether this striking similarity is a mere 
coincidence or not deserves further study, especially, in light of
the fact that the D-brane pair production at high-energy scattering
has been found for closed strings \cite{GHV} and hence cannot be a feature 
unique to open strings only.

\subsection*{Acknowledgement}
We would like to thank S. de Haro, Y. Kazama, H.B. Nielsen, G. Veneziano and 
T. Yoneya for invaluable discussions and helpful comments on high-energy 
scattering in string theory. SJR acknowledges warm hospitality of Theory 
Division at CERN, where the work is completed.


\begin{thebibliography}{9999}
\bibitem{witten}E. Witten, ``Bound States of Strings and $p$-Branes'', 
Nucl.~Phys.\ {\bf B460} (1996) 335, hep-th/9510135.
\bibitem{DH}M. R. Douglas and C. Hull, ``D-branes and the Noncommutative 
Torus,'' JHEP {\bf 02} (1998) 008, hep-th/9711165.
\bibitem{ishi}N. Ishibashi, ``$p$-branes from $(p-2)$-branes in the 
Bosonic String Theory'', Nucl.~Phys.\ {\bf B539} (1999) 107, hep-th/9804163; \\
``A Relation between Commutative and Noncommutative Descriptions 
of D-branes,'' hep-th/9909176.
\bibitem{KK}M. Kato and T. Kuroki, ``World Volume Noncommutativity 
versus Target Space Noncommutativity,'' JHEP {\bf 03} (1999) 012, 
hep-th/9902004.
\bibitem{SW}N. Seiberg and E. Witten, ``String Theory and Noncommutative 
Geometry,'' JHEP {\bf 09} (1999) 032, hep-th/9908142.
\bibitem{BFSS}T. Banks, W. Fishler, S.H. Shenker and L. Susskind, 
``M Theory as A Matrix Model: A Conjecture,'' Phys. Rev. {\bf D55} 
(1997) 5112, hep-th/9610043. 
\bibitem{CDS}A. Connes, M. R. Douglas and A. Schwarz, ``Noncommutative 
Geometry and Matrix Theory: Compactification on Tori,''
JHEP {\bf 02} (1998) 003, hep-th/9711162. 
\bibitem{IKKT}N. Ishibashi, H. Kawai, Y. Kitazawa and A. Tsuchiya, 
``A Large N Reduced Model as Superstring,'' Nucl.~Phys.\ {\bf B498} (1997) 
467, hep-th/9612115. 
\bibitem{AIKKT}H. Aoki, N. Ishibashi, S. Iso, H. Kawai Y. Kitazawa, T. Tada, 
``Noncommutative Yang-Mills in IIB Matrix Model,'' Nucl.Phys. {\bf B565} 
(2000) 176, hep-th/9908141; \\
N. Ishibashi, S. Iso, H. Kawai and Y. Kitazawa, ``Wilson Loops 
in Noncommutative Yang Mills,'' Nucl. Phys. {\bf B573} (2000) 573, 
hep-th/9910004;\\
S. Iso, H. Kawai, Y. Kitazawa, ``Bilocal Fields in Noncommutative 
Field Theory,'' Nucl. Phys. {\bf B576} (2000) 375, hep-th/0001027; \\
N. Ishibashi, S. Iso, H. Kawai, Y. Kitazawa, ``String Scale in Noncommutative 
Yang-Mills,''hep-th/0004038.  

\bibitem{SST}N. Seiberg, L. Susskind and N. Toumbas, ``Space/Time 
Non-Commutativity and Causality,'' JHEP {\bf 06} (2000) 044, hep-th/0005015. 

\bibitem{gomis} J. Gomis and T. Mehen, 
"Space-Time Noncommutative Field Theories and Unitarity,"
hep-th/0005129.

\bibitem{sjrey} S.-J. Rey, ``Open String Dynamics with Nonlocality in Time,'' 
to appear, SNUST-006005. 

\bibitem{SST2}N. Seiberg, L. Susskind and N. Toumbas, ``Strings in Background 
Electric Field, Space/Time Noncommutativity and A New Noncritical String 
Theory,'' JHEP {\bf 06} (2000) 021, hep-th/0005040.  

\bibitem{GMMS} R. Gopakumar, J. Maldacena, S. Minwalla and A. Strominger,
``S-Duality and Noncommutative Gauge Theory,'' JHEP {\bf 06} (2000) 036, 
hep-th/0005048.

\bibitem{BR}J. L. F. Barb\'{o}n and E. Rabinovici, ``Stringy Fuzziness 
as the Custodian of Time-Space Noncommutativity,'' hep-th/0005073. 

\bibitem{reyvonunge} S.-J. Rey and R. von Unge, ``S-Duality,
Noncritical Open String and Noncommutative Gauge Theory,'' 
SNUST-006002, hep-th/0007089.

\bibitem{stur}T. Yoneya, {\em  Duality and Indeterminacy Principle 
in String Theory} in ``Wandering in the Fields'', eds. K. Kawarabayashi 
and A. Ukawa (World Scientific, 1987), p. 419; {\em String Theory and 
Quantum Gravity} in ``Quantum String Theory'', eds. N. Kawamoto and 
T. Kugo (Springer,1988), p.23; \\
{\sl ibid} ``On The Interpretation of Minimal Length 
in String Theories'', Mod.~Phys.~Lett.\ {\bf A4} (1989) 1587.

\bibitem{yoneya}T. Yoneya, ``String Theory and Space-Time Uncertainty 
Principle,'' hep-th/0004074, and references therein. 

\bibitem{GM1} D. J. Gross and P. F. Mende, ``The High-Energy Behavior 
of String Scattering Amplitudes,'' Phys. Lett. {\bf B197} (1987) 129.

\bibitem{GM2} D. J. Gross and P. F. Mende, ``String Theory beyond 
The Planck Scale,'' Nucl. Phys. {\bf B303} (1988) 407. 

\bibitem{GMan}D. J. Gross and J. L. Ma\~{n}es, ``The High Energy Behavior 
of Open String Scattering, '' Nucl. Phys. {\bf B326} (1989) 73.  

\bibitem{gross}D. J. Gross, ``High-Energy Symmetries of String Theory,'' 
Phys. Rev. Lett. {\bf 60} (1988) 1229.

\bibitem{MO}P. F. Mende and H. Ooguri, ``Borel Summation of String 
Theory for Planck Scale Scattering,'' Nucl. Phys. {\bf B339} (1990) 641. 

\bibitem{GHV} S. B. Giddings, F. Hacquebord and H. Verlinde, 
"High-Energy Scattering and D-Pair Creation in Matrix String Theory,"
Nucl. Phys. {\bf B537} (1999) 260, hep-th/9804121.

\end{thebibliography}
\end{document}